\documentclass[twocolumn,english,aps,prb,superscriptaddress,nobinotes,amsmath,amssymb,floatfix]{revtex4-2}
\pdfoutput=1
\usepackage{amsfonts}
\usepackage{amsmath}
\usepackage{amssymb}
\usepackage[colorlinks=true,citecolor=blue,linkcolor=blue,breaklinks=true]{hyperref}
\usepackage{graphicx}
\usepackage{gensymb}

\usepackage{soul}
\usepackage{url}
\usepackage{dcolumn}
\usepackage{bm}
\usepackage{stmaryrd}
\usepackage{changes}
\usepackage{multirow}
\usepackage{rotating}
\usepackage{titlesec}
\usepackage{float}
\usepackage{tabularx}
\usepackage{textcomp}
\usepackage[flushleft]{threeparttable}
\usepackage{booktabs}
\usepackage{dcolumn}

\usepackage{titlesec}
\renewcommand{\thesection}{\arabic{section}}
\titleformat{\section}
{\normalfont\large\bfseries}{\thesection.}{1em}{}
\usepackage{afterpage}
\usepackage{soul}

\usepackage{hyperref}
\hypersetup{
    colorlinks=true,       
    linkcolor=blue,        
    citecolor=blue,        
    filecolor=magenta,     
    urlcolor=blue          
}

\newcommand{\figref}[1]{\mbox{Fig.~\ref{#1}}}

\newcommand{\tabref}[1]{\mbox{Table~\ref{#1}}}

\newcommand{\equalcontrib}{\thanks{These authors contributed equally.}}
\makeatletter
\def\@fnsymbol#1{\ensuremath{\ifcase#1\or *\or *\dagger\or \ddagger\or
   \mathsection\or \mathparagraph\or \|\or **\or \dagger\dagger
   \or \ddagger\ddagger \else\@ctrerr\fi}}
\makeatother

\newcommand{\IPHYS}{\affiliation{Institute of Physics, Swiss Federal Institute of Technology Lausanne (EPFL), Lausanne, Switzerland.}}
\newcommand{\EMI}{\affiliation{Institute of Electrical and Micro Engineering (IEM), EPFL, CH-1015 Lausanne, Switzerland}}
\newcommand{\QCENTER}{\affiliation{Center for Quantum Science and Engineering, Swiss Federal Institute of Technology Lausanne (EPFL), Lausanne, Switzerland.}}
\newcommand{\WW}{\affiliation{WithWave Co., Ltd., Quantum \& Space technologies, Yongin, Republic of Korea.}}

\begin{document}
	
	\title{Quantum-limited traveling-wave parametric amplifier based on DUV-defined planar structures}

	\author{Hao Li}\equalcontrib
	\IPHYS \EMI \QCENTER

	\author{Marco Scigliuzzo}\equalcontrib
	\email[]{marco.scigliuzzo@usherbrooke.ca}
	\IPHYS \EMI \QCENTER

	\author{Evgenii Guzovskii}\equalcontrib
	\IPHYS \EMI \QCENTER

	\author{Seog-Tae Han}
	\WW

	\author{Kyungho Han}
	\WW

	\author{Tobias~J.~Kippenberg}
	\email[]{tobias.kippenberg@epfl.ch}
	\IPHYS \EMI \QCENTER

	\begin{abstract}
	\end{abstract}

	\maketitle

		\textbf{The relentless scaling of classical microelectronics has been enabled by the precision and reproducibility of deep-ultraviolet (DUV) optical lithography. Implementing large-scale superconducting quantum processors will require cryogenic microwave components that follow a similarly scalable fabrication path. This need is particularly acute for high circuit-density devices such as traveling-wave parametric amplifiers (TWPAs), where recent implementations have demonstrated high gain, broad bandwidth, high saturation power, and near-quantum-limited noise, but trade-offs between footprint, insertion loss, and scalable integration remain. Here, we demonstrate a four-wave-mixing TWPA fabricated via a hybrid scheme that combines DUV-defined planar circuit elements with electron-beam-patterned Josephson junctions, constituting a first step toward fully scalable manufacturing. The device combines a compact footprint of 6.6\,mm$\times$ 6.6\,mm with broadband gain over the measured 3--11\,GHz span and an average 1\,dB compression point of -102\,dBm.  By using planar capacitors to reduce loss, it operates near the quantum limit,  with an average added noise of 0.4 photons above the standard quantum limit  in the 4 to 8\,GHz band. The phase-matching stopband remains narrow, with a bandwidth of 43\,MHz, consistent with resonator-frequency variation below 1\% and indicative of the circuit-parameter uniformity enabled by DUV lithography. These results show that DUV-defined planar elements can enable compact, low-loss, near-quantum-limited TWPAs and provide a promising route toward high-density cryogenic microwave hardware for large-scale quantum systems.}

High-fidelity single-shot qubit readout in superconducting quantum processors requires the amplification of weak microwave signals, often containing only a few photons, above the noise floor of room-temperature detectors \cite{clerk2010introduction}.
To maximize the signal-to-noise ratio, the first-stage amplifier should ideally operate at the standard quantum limit \cite{aumentado2020superconducting}.
While resonant Josephson parametric amplifiers (JPAs) have successfully fulfilled this role in many fundamental experiments \cite{castellanos2008amplification,vijay2011observation}, their resonant nature imposes a limited gain-bandwidth product \cite{roy2016introduction,mutus2014strong} and low saturation power \cite{eichler2014controlling}.
These constraints become a major bottleneck for frequency-multiplexed readout, which is essential for scaling superconducting quantum processors.

Traveling-wave parametric amplifiers (TWPAs) overcome these limitations by using nonlinear transmission lines \cite{cullen1958travelling,tien1958parametric} instead of a resonant cavity.
The most widely studied implementations rely on four-wave mixing in nonlinear transmission lines, with the nonlinearity provided either by Josephson junctions \cite{yaakobi2013parametric,o2014resonant,macklin2015near} or kinetic inductance \cite{ho2012wideband}.
More recently, three-wave mixing schemes have also been developed using DC-biased high-kinetic-inductance waveguides \cite{vissers2016low}, rf-SQUIDs \cite{zorin2017traveling,gaydamachenko2025rf}, and SNAIL arrays \cite{fadavi2023three, nilsson2024small}.
Continued efforts to improve TWPA performance have led to major advances, including nonlinearity engineering \cite{ranadive2022kerr,bell2026josephson} and the suppression of unwanted higher-order mixing processes \cite{Peng2022}.

Despite this rapid progress, compact integration remains a major challenge.
Recent TWPA implementations achieve very low added noise and insertion loss \cite{wang2025high}, but still rely on centimeter-scale chips.
While promising results can be achieved by eliminating resonant phase matching\cite{chang2025josephson}, resonantly phase-matched devices have the advantage of a narrower power-independent stopband which simplifies the design and operation of the TWPA. 
The integration of the phase matching resonators into a compact layout, however, remains a significant challenge.
In large-scale cryogenic quantum hardware, significant footprints limit the circuit density and complicate packaging by giving rise to spurious box modes.
Addressing these constraints is increasingly important as TWPAs are emerging not only as readout amplifiers, but also as useful sources of nonclassical microwave radiation \cite{esposito2022observation,perelshtein2022broadband,Qiu2023}.

Here, we present a compact resonantly phase-matched TWPA architecture that addresses this footprint--performance trade-off.
Using a hybrid wafer-level fabrication process, with DUV lithography for linear circuit elements and electron-beam lithography for the Josephson junctions, we realize a device with individual resonator footprints of $\sim100\,\mu$m$\times$170\,$\mu$m and a chip of 6.6\,mm$\times$ 6.6\,mm, much smaller than the field size of our DUV reticle.

			\begin{figure*}[t]
				\centering 
				\includegraphics[width=\textwidth]{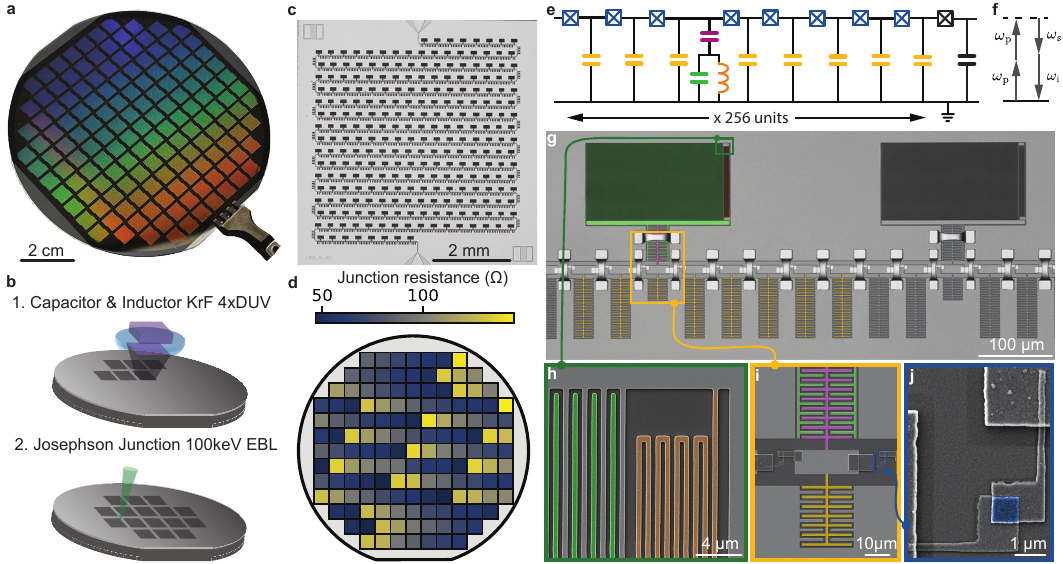} 
				\caption[Wafer-scale manufacturing of near quantum limited Josephson traveling-wave parametric amplifier]
				{
					\textbf{Wafer-scale manufacturing of near-quantum-limited Josephson traveling-wave parametric amplifier.} 
					\textbf{a}, the TWPA is realized on a 100\,mm wafer, hosting 145 individual chips.
					The colors arise from the interference of scattered light by circuit features comparable to the optical wavelength.
					\textbf{b}, the core fabrication process relies on two primary lithography steps.
					Deep-ultraviolet (DUV) lithography (utilizing a 4$\times$ magnification mask) defines the waveguide capacitors, resonator capacitors, and meandered inductors.
					Electron-beam lithography (EBL) is subsequently used to define the Josephson junctions (JJs).
					The process is completed with junction patches and air-bridges to connect the ground planes (not shown).
					\textbf{c},  optical micrograph of a single TWPA chip post-dicing.
					\textbf{d}, room-temperature resistance mapping of single JJs, measured on test structures comprising 10- or 11-junction arrays.
					Junction dimensions are swept across nine discrete values.
					 \textbf{e}, equivalent circuit schematic of a single TWPA unit cell, consisting of eight JJs and a phase-matching resonator.
					This unit cell is cascaded 256 times along the transmission line.
					 \textbf{f}, amplification relies on a four-wave mixing process, wherein two pump photons at frequency $\omega_{\rm p}$ are converted into a signal photon at $\omega_{\rm s}$ and an idler photon at $\omega_{\rm i}$.
					\textbf{g}, false-color scanning electron microscope (SEM) micrograph of a unit cell.
					Highlighted components include the phase-matching resonator capacitor (green) and inductor (orange), detailed further in \textbf{h}.
					\textbf{i}, Close-up of the resonator coupling capacitor (purple) and the capacitance to ground (yellow).
					\textbf{j}, SEM micrograph detailing the Josephson junctions (blue).
				}
				\label{fig:concept} 
			\end{figure*}

 Importantly, this approach preserves sub-dB insertion loss, as required for compatibility with high-coherence superconducting qubit platforms.
 Moreover, the observation of a narrow phase-matching stopband is consistent with small resonator-frequency variation, estimated to be below 1\%.
 The amplifier exhibits broadband gain over the measured 3--11\,GHz frequency span, while maintaining an average 1\,dB compression point of -102\,dBm between 4 and 8\,GHz.
 Our TWPA adds on average 0.4 photons per second per Hz above the quantum limit over the same frequency range, positioning this planar architecture as a promising route toward scalable multiplexed readout.

				\subsection*{Implementation of compact planar TWPA}
			We fabricate the TWPAs on a 100\,mm silicon wafer as shown in \figref{fig:concept}a using sputtered niobium for the ground plane, and evaporated aluminum for the Josephson junctions and the air bridges.
			The fabrication begins with patterning the linear components (inductors and capacitors) on the 150\,nm thick Niobium ground plane with DUV lithography (248\,nm wavelength KrF laser, with a 4-times reduction of the mask image). 
			The lithography takes less than 10 minutes per wafer, including spin coating, post exposure bake and development of the photoresist.
			The ground plane etching is then performed using inductively coupled plasma reactive ion etching with chlorine chemistry. 
			Electron-beam lithography (EBL) and shadow electron-beam evaporation of aluminum are subsequently used to define the Josephson junctions (JJs) (see \figref{fig:concept}b).
			The process is completed with junction patches and air-bridges to connect the ground planes. 
			An optical micrograph of a single TWPA chip post-dicing is shown in \figref{fig:concept}c. 
			The chip dimensions are $6.6\times 6.6$\,mm$^2$, which facilitates compact packaging and prevents spurious chip modes within the operating band. 
			The junction area was varied across 9 different values with a relative increment of 20\%, as shown in the map of the room-temperature resistance of single JJs, reported in \figref{fig:concept}d.
			This mitigates the run-to-run variability in the critical current density of the junctions due to oxidation conditions and ensures devices with the desired JJ critical current are present on the wafer.

			\begin{figure*}[t]
				\centering 
				\includegraphics[width=\textwidth]{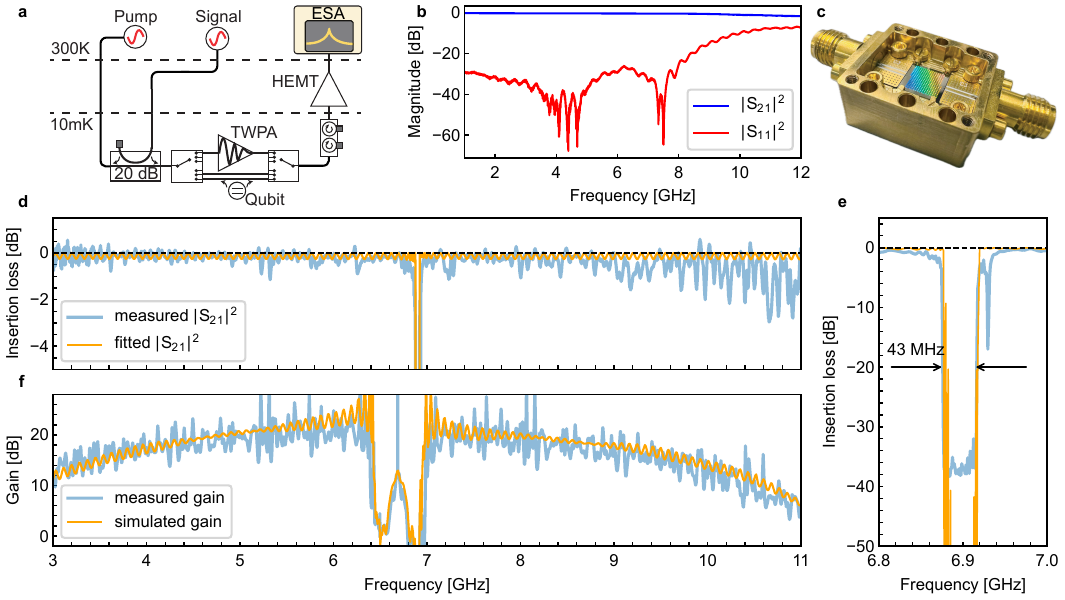} 
				\caption[Amplifier performance]
				{
					\textbf{Performance characterization of the traveling-wave parametric amplifier.} 
					\textbf{a}, Simplified schematic of the measurement setup.
					The TWPA is mounted on the mixing chamber stage of a dilution refrigerator, with the signal and the pump tone injected through a directional coupler.
					A bypass coaxial cable and a qubit directly coupled to the waveguide allow for direct transmission measurements without the amplifier and power calibration, respectively.
					\textbf{b}, Transmittance $|S_{21}|^2$ and reflectance $|S_{11}|^2$ of the sample packaging at room temperature for a gold straight CPW.
					\textbf{c}, Packaged TWPA without the box's lid.
					\textbf{d}, Insertion loss of the TWPA without the pump, measured by comparing the transmission through the TWPA with the bypass cable.
					\textbf{e}, A detailed scan of the phase-matching stopband, highlighting the narrow 43\,MHz rejection region.
					The continuous line represents a theoretical fit to the propagation in the TWPA line.
					\textbf{f}, Gain profile measured by comparing the TWPA output with the transmission through the bypass coaxial cable.
					Around the pump frequency at 6.688\,GHz, two regions of no gain are visible due to the phase-matching stopband.
					The continuous orange line represents the simulated gain using extracted parameters.
					Crucially, the ripples observed in the measured gain are consistent with  impedance mismatch and package resonances.
				}
				\label{fig:gain} 
			\end{figure*}

			The TWPA is implemented as a nonlinear transmission line of 256 cascaded unit cells, each comprising eight series-connected Josephson junctions with planar shunt capacitors and a single capacitively coupled phase-matching resonator, as illustrated by the equivalent circuit schematic in \figref{fig:concept}e.
			The amplification relies on a four-wave mixing process, wherein two pump photons at the frequency $\omega_{\rm p}$ are converted into a signal photon at $\omega_{\rm s}$ and an idler photon at $\omega_{\rm i}$, as shown in \figref{fig:concept}f. 
			A false-color scanning electron microscope (SEM) micrograph of a unit cell is shown in \figref{fig:concept}g, where the phase-matching resonator's capacitor (green) and inductor (orange) are highlighted.
			A close-up of the resonator coupling capacitor (purple) and the capacitance to ground (yellow) are shown in \figref{fig:concept}i.
			Finally, a SEM micrograph detailing the Josephson junctions (blue) is shown in \figref{fig:concept}j.

			To achieve a compact footprint and avoid the appearance of chip modes we minimize the pitch of the features of the resonator's inductor and capacitor to 800\,nm, and to 2\,$\mu$m for the coupling capacitors.
			The junctions have a critical dimension of 250\,nm for the contact pads however the dimensions of the junction area is close to $1\times1\,\mu$m$^2$.  

		\subsection*{Amplification performance}
			Internal reflections and impedance mismatch between the TWPA and the 50\,$\Omega$ environment can lead to ripples in the gain profile\cite{renberg2024peripheral}, which in turn cause ripples in the added noise.
			To facilitate the interface between the 50\,$\Omega$ coaxial cable and the on-chip transmission line, we realize a package based on small PCBs with short 50\,$\Omega$ transmission lines as shown in \figref{fig:gain}c. 
			To shift the cavity and chip modes of packaged assembly beyond the TWPA's operating frequency range, a vacuum pocket with optimized dimensions is added beneath the chip.
			Furthermore, custom-designed  connectors are employed to enhance the grounding interface between the connectors and the PCBs.

			The sample packaging is characterized at room temperature by measuring the transmittance and reflectance of the package containing a straight gold coplanar waveguide replacing the TWPA chip, reported in \figref{fig:gain}b.
			The reflectance of the package is well below -20\,dB up to 8\,GHz, which is important for achieving low ripple in the gain and added noise of the amplifier. 
			

			\begin{figure*}[t]
				\centering 
			    \includegraphics[width=\textwidth]{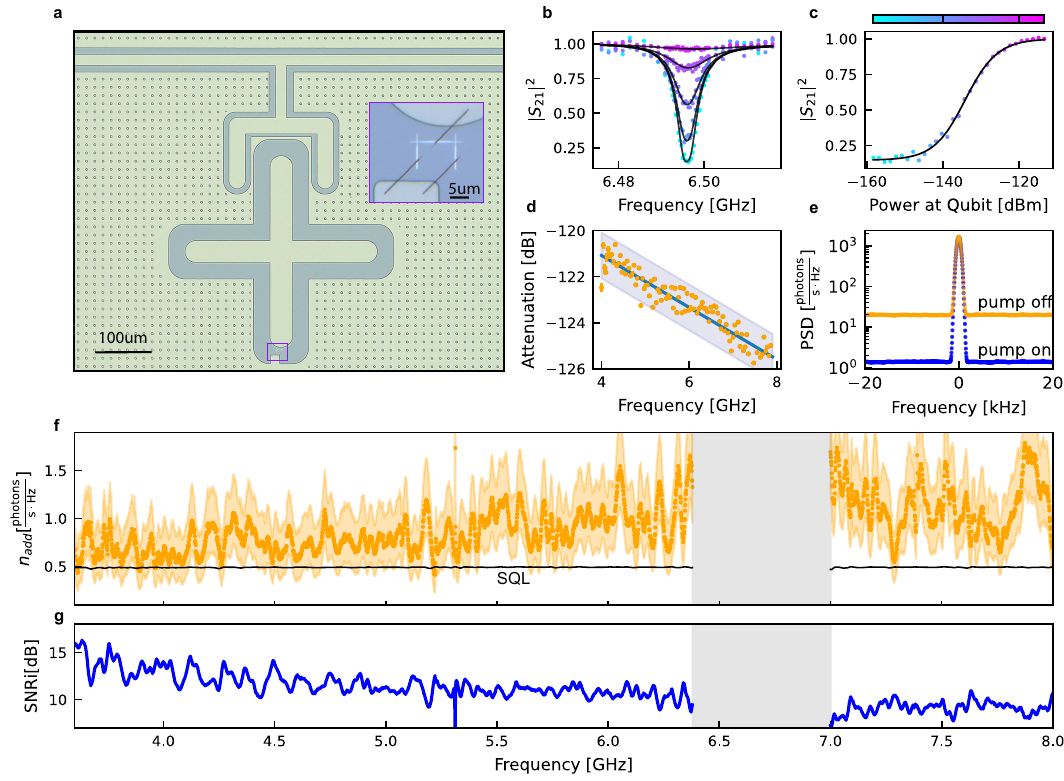} 
			    \caption[Noise performance of the TWPA]
			    {
			    	\textbf{Noise performance of the near-quantum-limited Josephson junctions based TWPA.}
			    	\textbf{a}, Optical micrograph of the frequency-tunable   waveguide coupled transmon used for signal and pump power calibration.
					The inset shows a magnified view of the SQUID.
					\textbf{b}, Transmittance ($|S_{21}|^2$) of the qubit feedline at varying driving powers.
					At low power, the resonance indicates the qubit is strongly overcoupled to the feedline.
					Solid lines represent a global complex fit.
					\textbf{c}, Power dependence of the feedline transmittance on resonance with the qubit.
					The solid line indicates the fit.
					\textbf{d}, Signal line attenuation extracted across the 4 to 8 GHz qubit tuning range, utilizing the calibrated power at the qubit.
					The attenuation exhibits a linear frequency dependence in dB (solid line indicates linear fit, shaded blue area indicates $\pm$1\,dB deviation from the linear fit).
					\textbf{e}, Calibrated power spectral density (PSD) of a 5 GHz signal normalized to the input of the TWPA, measured with the parametric pump turned on and off.
					\textbf{f}, Calibrated added noise ($n_{add}$) in units of photons (solid orange points), with the shaded region representing the measurement uncertainty.
					The standard quantum limit (SQL, solid black line) accounts for the frequency-dependent parametric gain.
					The greyed-out band indicates the frequency range where the phase-matching stopband on either the signal or idler inhibits amplification.
					\textbf{g}, Signal-to-noise ratio improvement (SNRi), determined by the difference between the SNR measured through the TWPA and that
					measured through the bypass path.
				}
			    \label{fig:noise} 
			\end{figure*}

			\begin{figure*}[t]
				\centering 
				\includegraphics[width=\textwidth]{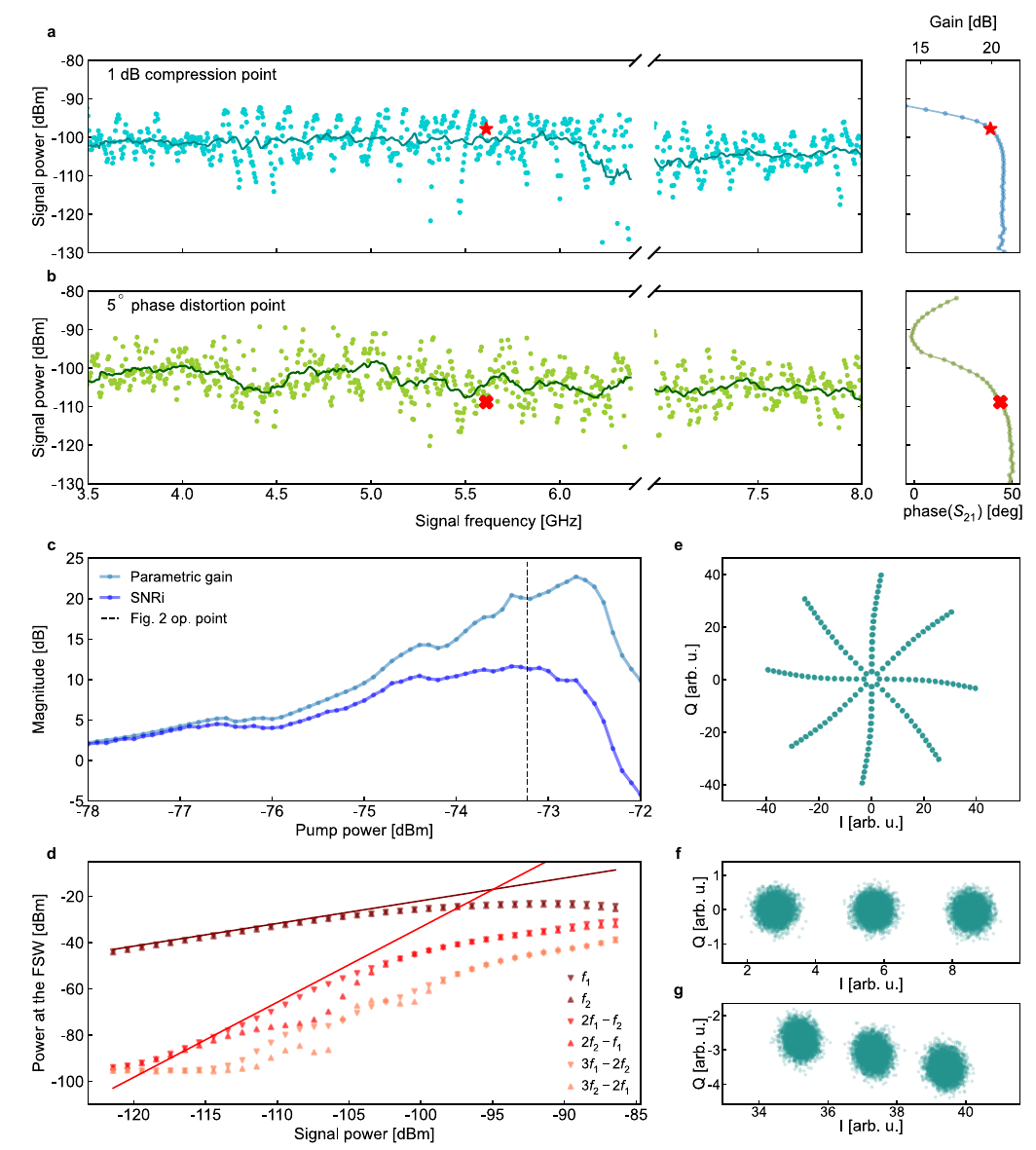} 
				\caption[Power handling]
				{
					\textbf{Power handling of the amplifier}					
					\textbf{a}, Amplifier 1dB compression point for various signal frequencies within the operational range of frequencies
					\textbf{b}, 5$^o$ phase distortion point, i.e. power at which the phase of the amplified signal is shifted 5$^o$ with respect to the phase of a small signal at the same frequency.
					\textbf{c}, Small signal gain and signal-to-noise ratio improvement versus pump power at the input of the amplifier.
					\textbf{d}, Measured powers of different intermodulation products of two tones centered at 5 GHz and separated by 5 MHz versus their power at the input of the amplifier.
					\textbf{e}, IQ constellation diagram obtained from repeated fixed-duration pulse-readout measurements with the TWPA pump on. The pulse amplitude is swept linearly for eight different phases.
					\textbf{f}, Zoom-in on the three lowest amplitudes for one phase. The IQ blobs remain circular for the signal power below 1 dB compression point.
					\textbf{g}, Zoom-in on the three highest amplitudes for the same phase as in panel \textbf{f}. TWPA gain compression reduces the spacing between neighboring IQ blobs and deforms their circular shape.
				}
				\label{fig:power_handling} 
			\end{figure*}

			In \figref{fig:gain}d we report the insertion loss of the TWPA without the pump, measured by comparing the transmission through the TWPA to the transmission of a bypass cable.
			The insertion loss is below 1\,dB across the 3 to 8\,GHz band, which is crucial for achieving quantum-limited noise performance\cite{Peng2022, Qiu2023}. 
			In \figref{fig:gain}e we report a detailed scan of the phase-matching stopband, highlighting the narrow 43\,MHz rejection region.
			In both panels, the continuous line represents a theoretical calculation of the transmission through the TWPA using ABCD-matrix method with parameters summarized in \tabref{tab:twpa_parameters}.
			These parameters are obtained by fitting the unwrapped pump-off transmission phase referenced to the bypass line.
			The model also includes phase-matching-resonator frequency disorder, described by a Gaussian distribution, as detailed in Appendix~\ref{app:TWPA parameters}.




			In \figref{fig:gain}f we report the gain measured by comparing the output of the TWPA with the transmission through a bypass coaxial cable.
			The gain exceeds 10\,dB over most of the 3 to 11\,GHz frequency span, except for a 600\,MHz band around the pump frequency at 6.688\,GHz where the propagation of either the signal or the idler is prohibited by the phase-matching stopband. 
			The measured gain profile exhibits ripples of several decibels, which we attribute to the impedance mismatch between the TWPA and the $50\,\Omega$ environment, compounded by additional reflections from peripheral circuitry.


			The continuous orange line represents the gain calculated using a harmonic-balance circuit simulation \cite{OBrien_JosephsonCircuits_jl,peng2022xparameter} with the circuit parameters listed in \tabref{tab:twpa_parameters}. Additionally, as mentioned above,  the resonance frequencies of the phase-matching resonators were sampled from a normal distribution while keeping their impedance fixed, to mimic random fabrication-induced variability. The mean and standard deviation of this distribution were chosen to reproduce the measured center frequency and linewidth of the phase-matching stopband, whose width cannot be explained by the external coupling rate of nominally identical resonators alone.

			The gain of the amplifier can be considered in the large gain limit as $G>10$\,dB in almost the entire measured frequency span. 
			Therefore it is possible to compare the measured added noise of the amplifier with the standard quantum limit (SQL) of half a photon per second per Hz for a phase-insensitive amplifier.

			In order to precisely calibrate the attenuation of the signal and the pump lines we use a frequency-tunable transmon qubit overcoupled to a waveguide in a notch configuration (see \figref{fig:noise}a and inset).
			The frequency of the transmon is tuned with a superconducting coil biased with a DC current filtered at the 4K stage.
			The transmittance around the qubit resonance is measured at varying driving powers, and reported in \figref{fig:noise}b.
			At high probe powers the qubit is completely saturated while at lower powers the qubit response follows a Lorentzian.
			At resonance this gives the characteristic shape reported in \figref{fig:noise}c.
			These transmittances are globally fit as a function of probe power and frequency (solid black line), giving a primary estimator for the attenuation between the qubit reference plane and the output reference plane of the VNA (see Appendix \ref{app: power}). 
			The extracted attenuation values within the qubit tunability range of 4 to 8 GHz are reported in \figref{fig:noise}d.
			The attenuation exhibits a linear frequency dependence in logarithmic scale, which is expected for coaxial cables used in the measurement setup.
			We fit the extracted attenuation with a linear function (solid line) and use the fitting function to calibrate the power spectra measured with a spectrum analyzer in units of quanta per s per Hz.
			Measuring the power spectra with and without parametric pump (see \figref{fig:noise}e where we report the power spectral density (PSD) with the pump on and pump off for the signal frequency of 5 GHz normalized to the input of the TWPA) we extract the added noise referred to the input of the TWPA ($n_{\rm{add}}$), which is plotted in \figref{fig:noise}f, using the following equation:
			\begin{equation}
				\label{eq:added_noise}
				n^{\rm{T}}_{\rm{add}} = \left(\frac{N_{\rm{on}} - N_{\rm{off}}}{G_{\rm{sys}}} + \frac{1}{2}\right) \frac{1}{G_{\rm{T}}} - \frac{1}{2},
			\end{equation} 
			where $N_{\rm{on}}$($N_{\rm{off}}$) is the equivalent noise PSD in units of photons per second per Hz with the TWPA pump turned on (off),extracted by averaging the flat noise floor outside the coherent signal peak;
			$G_{\rm{sys}}$ is the gain of the readout chain after the TWPA, and $G_{\rm{T}}$ is the parametric gain of the TWPA.

			The amplifier achieves an average added noise of 0.4 photons per second per Hz above the standard quantum limit within the 3.5 to 8 GHz range. The greyed-out band indicates the frequency range where the phase-matching stopband prohibits the propagation of either the signal or the idler, and is not taken into account when calculating the average added noise. 
			Finally, we determine the signal-to-noise ratio improvement (SNRi) by subtracting the baseline SNR through a bypass cable from the amplified signal SNR (see \figref{fig:noise}g).

		\subsection*{Power handling}

Intermodulation products can severely degrade amplifier performance, particularly in frequency-multiplexed qubit readout schemes \cite{remm2023intermodulation}. To establish the operational limits of our device, we systematically characterized its power handling capabilities.

The small-signal gain increases with pump power, reaching a maximum of approximately 20\,dB at an incident pump power of -73\,dBm (see \figref{fig:power_handling}c).
However, the signal-to-noise ratio improvement (SNRi) peaks at a lower pump power.
This divergence between maximum gain and maximum SNRi indicates the onset of parametric oscillation, as the amplifier is pushed toward saturation.
Consequently, the optimal operating point for the TWPA is chosen to strictly maximize SNRi rather than absolute gain.
In \figref{fig:gain}, we operate at this SNRi-optimized point, indicated by the dashed black line. 

In \figref{fig:power_handling}d we report the output power of various intermodulation products generated by two tones centered at 5\,GHz and separated by 5\,MHz, plotted against their input power.
The ratio of the signal to the intermodulation products decreases rapidly with increasing input power.
At low input powers, the signal exceeds the intermodulation products by 40\,dB.
However, near the amplifier's saturation point, this margin narrows to 10\,dB, making intermodulation distortion a non-negligible factor for readout fidelity.
Extrapolating this trend yields a third-order intercept point (IP3) of approximately -90\,dBm, which is 10\,dB higher than the measured 1\,dB compression point of -100\,dBm at this frequency.

In \figref{fig:power_handling}a we report the 1\,dB compression point across the operational frequency range.
Additionally, \figref{fig:power_handling}b shows the $5^\circ$ phase distortion point, defined as the input power that induces a $5^\circ$ phase shift in the amplified signal (relative to the VNA local oscillator) compared to the small-signal phase response.
Both the 1\,dB compression point and the $5^\circ$ phase distortion point remain close to -100\,dBm across all measured frequencies. 
This provides a substantial power margin for typical dispersive readout scenarios, where the signal power arriving from a single resonator is roughly -130\,dBm, comfortably enabling the simultaneous readout of multiple qubits before saturation effects occur.

To further illustrate the impact of the amplifier saturation on dispersive readout, we measured the quadrature (I and Q) response to fixed-duration microwave pulses with the TWPA pump on (Fig. 4e--g).
In the low-power regime, the IQ-cluster centroids scale linearly with the applied pulse amplitude, while their angle in the IQ plane remains unchanged. At higher input powers, however, TWPA gain compression reduces the spacing between neighboring IQ blobs, while the nonlinear phase response rotates the amplified signal in the IQ plane and deforms the circular distributions.
In a multiplexed dispersive-readout setting, where the total signal power at the amplifier input increases with the number of simultaneously measured resonators, such amplitude- and phase-dependent distortion can modify the calibrated readout axes and decision thresholds.
These measurements therefore provide a direct IQ-plane visualization of the dynamic-range limit relevant for scaling simultaneous qubit readout.

\subsection*{Discussion and Conclusion}

Our TWPA demonstrates near-quantum-limited performance over an instantaneous bandwidth of approximately 3\,GHz, spanning from 3.5 to 6.5\,GHz. 
The residual ripple in the added-noise spectrum follows the oscillations in the gain profile, indicating that both are likely governed by the same underlying impedance mismatches within the device and the setup, which introduce frequency-dependent reflections.
To assess the role of the external impedance environment, we performed an additional diagnostic measurement with two $10\,\mathrm{dB}$ cryogenic attenuators placed immediately before and after the TWPA package, as described in Appendix~\ref{app:gain ripple}. The attenuated configuration showed a reduced gain-ripple amplitude but did not lead to a systematic increase in the maximum stable gain. This suggests that the parametric oscillation threshold is not primarily set by the external impedance environment, but rather by disorder internal to the nonlinear transmission line. A likely source of such disorder is cell-to-cell variation in the Josephson-junction critical currents, which aligns with the geometrical variation of the junction areas across different positions on the chip.
This identifies junction fabrication as the key target for further improvement, specifically through the optimization of the lithography process and the transition to a fully DUV-defined architecture \cite{macklin2015near,van2024advanced,ke2026scaffold}.

The measured 1\,dB compression point, averages approximately -102\,dBm, is compatible with circuit-QED multiplexed dispersive readout with single readout tone power around -130\,dBm.
Relative to the previous implementations of resonantly phase-matched TWPAs, including the device reported in \cite{macklin2015near} or photonic crystal TWPA \cite{planat2020photonic}, our amplifier achieves a similar gain and bandwidth with substantially lower insertion loss.
This represents a key practical advantage over conventional dielectric-based TWPAs. While high off-state losses typically degrade the measurement SNR, our device maintains a remarkably low insertion loss when unpumped, ensuring efficient qubit readout even in the off-state.
Compared with more recent low-loss resonant phase matching TWPA implementations \cite{wang2025high} on $5\times40$\,mm$^2$ chips, the present device is competitive in gain and bandwidth while offering a five times smaller footprint of $6.6\times6.6$\,mm$^2$. 

The central advance of this work is therefore not a single extreme performance metric, but rather the combination of near-quantum-limited operation, low insertion loss, compact footprint, and compatibility with wafer-level planar fabrication.
These results support compact resonantly phase-matched TWPAs as a credible route toward scalable cryogenic microwave hardware and frequency-multiplexed readout.
Further progress should be possible by extending the fabrication flow toward deep-UV-defined Josephson junctions, in line with processes already developed at the industrial level for superconducting qubit technologies.

{\footnotesize
	\subsection*{Author Contributions}
 		M.S., H.L. and E.G. designed the experiment. H.L. designed the sample with the input from M.S. . H.L. developed the fabrication process with input from E.G. and M.S. . H.L. fabricated the samples. E.G. and H.L. took the measurements reported in the manuscript. H.L., E.G. and M.S. analyzed the data. S.H. and K.H. designed, assembled and tested the sample packaging. T.J.K. supervised the project. E.G., M.S. and H.L. wrote the manuscript with input from all authors.

 	\subsection*{Acknowledgement}
		We thank Mahdi Chegnizadeh for realizing gold CPWs and measuring packaging scattering and Shingo Kono for insightful discussions.
		This work was supported by Innosuisse via the International project 119.473 INT-ENG (Q-LAEP), by the EU Horizon Europe research and innovation programme under grant No. 101248749 (QuAMP) and by EPFL via the QTNet initiative.
		M.S. acknowledges support from the EPFL Center for Quantum Science and Engineering postdoctoral fellowship. This work was also supported by the Korea Institute for Advancement of Technology (KIAT) grant funded by the Korea Government (MOTIE) (Quantum LAEP: Quantum Limited Amplifiers with Engineered Pump, Project No. P0028266). The device was fabricated in the Center of MicroNanoTechnology (CMi) at EPFL, and the packaging was realized by WithWave.
}
\appendix
	\vspace{0.4in}

	\section{Simulation and layout design}

		\subsection{Transmission line Design}

			To realize compact, low-loss shunt capacitance in the nonlinear transmission line, we use planar interdigitated capacitors with a 2-µm finger pitch and a 1-µm finger gap. 
			To accurately evaluate the unit-cell capacitance including the parasitic contributions from the airbridges, we performed S parameters simulations of 10 unit cells using Sonnet. In this model, the airbridges were represented by an additional metal layer matching their exact geometry 2\,$\mu$m above the transmission line, connected to the ground plane through vias. 


			To model the Josephson junctions, we inserted ideal lumped-element inductors along the simulated signal line and swept their inductance in steps of 5\,pH. For each swept value, we extracted the corresponding scattering parameters. 
			We then performed a global two-dimensional fit of the unwrapped phase as a function of both frequency and inductance, using a standard transmission-line model.
			To ensure self-consistency across the sweep, the fit was constrained by requiring the unit-cell inductance of successive datasets to differ by exactly the simulated 5\,pH step.
			This global fitting procedure yielded the total capacitance and unit-cell inductance, from which we successfully extracted the intrinsic stray linear inductance of the circuit.

			\begin{figure}[h]
				\centering 
			    \includegraphics[width=\columnwidth]{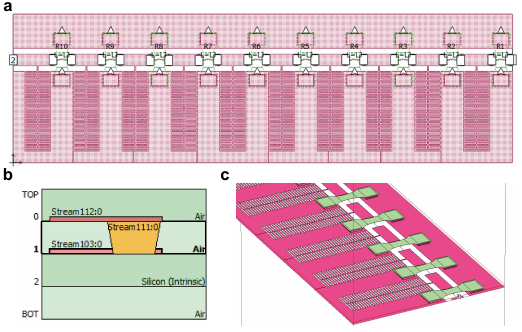} 
			    \caption[Sonnet simulation of transmission line]
			    {
			    	\textbf{Sonnet simulation of the transmission line with 10 unit cells. a,} Top-down layout of the 10-unit-cell model, with ideal lumped-element inductors inserted along the signal line to simulate the Josephson junctions. \textbf{b}, Cross-sectional stackup detailing the  silicon substrate, primary planar circuit layer,  the top airbridge metal layer  connected to the circuit layer by vias. \textbf{c},  Three-dimensional view highlighting the airbridges and vias. The out-of-plane features appear compressed because the lateral circuit dimensions are much larger than the 2\,$\mu$m vertical air gap.
				}
			    \label{fig:Sonnet_TL} 
		\end{figure}

		\subsection{Design of phase matching resonator}
			The phase-matching resonators consisted of interdigitated capacitors and meandered inductors with a nominal critical width of 400\,nm. To account for lithography and etching bias, the simulations used the measured fabricated dimensions: a capacitor-finger width of 420\,nm and an inter-finger gap of 380\,nm. We estimated the resonance frequency with 2.5D electromagnetic simulations in Sonnet. An excitation port was placed at the pad connecting the meandered inductor, allowing extraction of the effective one-port inductance. The meandered inductor and interdigitated capacitor form a parallel LC resonator, producing a clear resonance feature.

			The initial simulations treated the metal layers as perfect electric conductors. Experimentally, however, the measured stopband frequency was approximately 1 \,GHz lower than this idealized simulated value. We ruled out substrate overetching as the dominant origin, because etching into silicon would reduce the effective dielectric constant and therefore increase, rather than decrease, the resonator frequency. We instead attribute the downward shift to the kinetic inductance of the niobium film. Since the meandered-inductor traces are only approximately 400 \,nm wide, the kinetic inductance makes an appreciable contribution to the total resonator inductance. We therefore included a sheet kinetic inductance of 0.077 $pH/\square$ in the simulation. The simulated one-port inductance was then fitted to a shunted-LC model to extract the effective inductance and capacitance of the phase-matching resonators.

			\begin{figure}[h]
				\centering 
			    \includegraphics[width=\columnwidth]{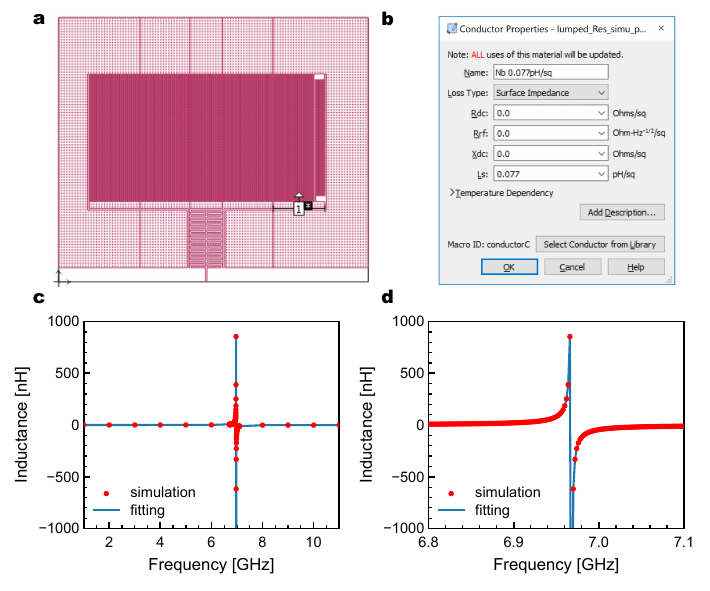} 
			    \caption[Sonnet simulation of phase matching resoantor]
			    {
			    	\textbf{Sonnet simulation of the phase-matching resonator.}  \textbf{a}, Layout of the interdigitated capacitor and meandered inductor. An excitation port is connected to the meandered inductor to extract the one-port inductance. \textbf{b}, Conductor properties used to assign a kinetic inductance of 0.077 $pH/\square$ through the surface impedance, which is required to model the narrow traces and match the measured stopband frequency. \textbf{c}, Extracted one-port inductance versus frequency. The simulated data (red markers) are fitted to a shunted LC model (blue line), revealing a resonance. \textbf{d}, Enlarged view around the resonator frequency.
				}
			    \label{fig:Sonnet_phase_matching_resonator} 
		\end{figure}
			
		\subsection{TWPA parameters} 
		\label{app:TWPA parameters}
		
		The device parameters used in this work were determined by combining room-temperature resistance measurements, electromagnetic simulations, and fits to the experimental pump-off transmission data. The junction critical current $I_c$ was estimated independently from the room-temperature normal-state resistance of co-fabricated test junction arrays (see top left and bottom right corners in Fig. \ref{fig:concept}c). The intrinsic stray linear inductance of the signal line, $L_s$, was extracted from the Sonnet transmission-line simulations described above. The initial unit-cell capacitance to ground $C_g$ was estimated from the measured phase velocity, as the Sonnet-extracted value ($\sim 48\,\mathrm{fF}$) was incompatible with the measured phase velocity and Junction normal-state resistance. We attribute this discrepancy to the limited mesh density available when simulating such a large physical footprint. 
		The initial junction capacitance $C_J$ was estimated from the junction area assuming $40\mathrm{fF}/\mu\mathrm{m}^2$. Resonator-frequency disorder was modeled as a Gaussian distribution in frequency with a fixed common impedance. With $I_c$ and $L_s$ fixed to their independently determined values,the remaining physical parameters were then refined by fitting the unwrapped pump-off signal-transmission phase relative to the bypass line in the same microwave switch banks. The resulting fitted parameters are summarized in Table \ref{tab:twpa_parameters}.

	\begin{table}[h]
    \centering
    \caption{estimated TWPA parameters}
    \label{tab:twpa_parameters}
    \begin{tabular}{lcc}
        \hline
        Parameter & Value & Unit \\
        \hline
        $I_c$ & 3.36 & $\mu\mathrm{A}$ \\
        $C_g$ & 70 & fF \\
        $L_s$ & 9.73 & pH \\
        $C_c$ & 26 & fF \\
        $C_J$ & 40 & fF \\
        $L_r$ & 0.41 & nH \\
        $C_r$ & 1277.92 & fF \\
        $\sigma_{f_r}$ & 7.96 & MHz \\
        \hline
    \end{tabular}
\end{table}
		
		\subsection{Design of waveguide QED device}
		The waveguide QED device was designed as a calibration device for extracting the microwave attenuation at the TWPA input. 
		For the electromagnetic design, the Josephson junction was replaced by a lumped linear inductor. This approximation removes the qubit nonlinearity and turns the circuit into a linear resonator, while preserving the microwave admittance that sets the coupling between the resonator and the feedline. We simulated the complex feedline transmission $S_{21}(\omega)$ around the resonator mode and fitted it to a notch-resonator response. 
		This fit provides the resonator--feedline coupling used to choose an overcoupled geometry for power calibration. 

		\subsection{Simulation of the TWPA gain}
			The gain of our TWPA was simulated using harmonic-balance method with the JosephsonCircuits.jl library \cite{OBrien_JosephsonCircuits_jl}. To accurately capture the physical behavior of the fabricated device, our model explicitly incorporated the intrinsic stray linear inductance of the transmission line. Furthermore, we accounted for fabrication-induced spatial disorder by applying parameter scattering to the phase-matching resonator frequencies assuming the impedance of the resonators are not changed. The frequencies of each resonator were sampled from a normal distribution utilizing the mean and standard deviation extracted directly from fits to the experimental transmission measurements. We swept the signal frequency from 3 to 11 GHz in 10-MHz steps to precisely map the gain profile and capture potential ripples. To obtain sufficient accuracy in our nonlinear analysis, we used the harmonic-balance method with 20 harmonics of the pump frequency $\omega_p$ to solve the large-signal nonlinear circuit response, and 10 signal/idler mixing products for the linearized small-signal analysis. This harmonic truncation ensures that the dominant four-wave mixing processes and relevant higher-order sidebands are adequately captured for an accurate prediction of the parametric gain.

	\section{Detailed measurement setup}

\begin{figure}[h]
				\centering 
			    \includegraphics[width=\columnwidth]{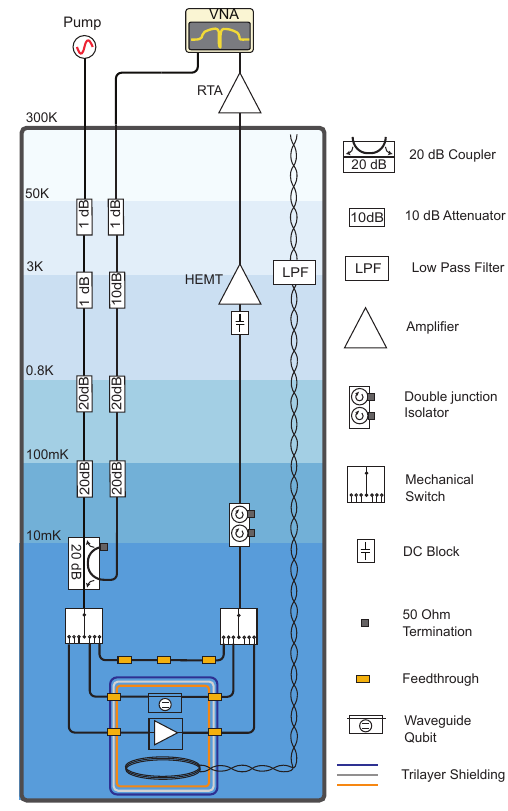} 
			    \caption[Measurement setup]
			    {
			    	\textbf{Measurement setup.} Schematic of the measurement setup used to characterize the TWPA. 
				}
			    \label{fig:setup} 
		\end{figure}

	After fabrication, the TWPA chip was fixed in the package and wirebonded with approximately 20 wires per ground side and five wires for the signal line. The package was then mechanically and thermally anchored to the mixing-chamber stage of a dilution refrigerator with a base temperature of 14\,mK.  The measurement setup is shown in \figref{fig:setup}.

		The TWPA, waveguide-QED calibration device, and bypass coaxial cable were all connected between the same pair of cryogenic microwave switches. The TWPA and qubit samples were enclosed in a three-layer magnetic shield consisting of an inner oxygen-free copper layer, a middle aluminium layer, and an outer mu-metal layer. Inside the enclosure, a superconducting coil driven by a DC current through twisted-pair superconducting lines was used to tune the qubit frequency. To ensure accurate insertion-loss and added-noise calibration, the coaxial cable sections outside the devices under test were made nominally identical in length for the three switch-selected paths. Because the TWPA and waveguide-QED paths passed through the shield, each included two microwave feedthroughs; identical feedthroughs were also added to the bypass coaxial path to keep the three paths as similar as possible.

		The signal and pump tones were combined at the mixing-chamber stage using a directional coupler (Pasternack PE2211-20). The TWPA pump was generated by a microwave source (Rohde $\&$ Schwarz SMA100B) with very low phase noise and clean spectrum with minimal sidebands. For the insertion-loss, gain, signal-to-noise-ratio-improvement, and intermodulation-product measurements, the signal tone was generated by a microwave source (AnaPico APMS12G). For the power calibration, 1 dB compression point, and 5 degree phase distortion measurements, the probe signal was generated by a vector network analyzer (Rohde $\&$ Schwarz ZNB20). 
		The output signal was routed through a 4--12 GHz isolator  (NF-ISISC4$\_$12A), a DC block and a HEMT amplifier (LNF-LNC4$\_$16A)  and then further amplified by a room-temperature amplfier (Mini-Circuits ZVA-183-S+).The output was then measured with either a vector network analyzer (Rohde $\&$ Schwarz ZNB20) or a spectrum analyzer (Rohde $\&$ Schwarz FSW26), depending on the measurement. For the constellation-diagram measurements, the microwave pulses were generated and acquired using a Zurich Instruments SHFQC4.

	\section{Power calibration}
	\label{app: power}
\begin{figure}[h]
				\centering 
			   \includegraphics[width=\columnwidth]{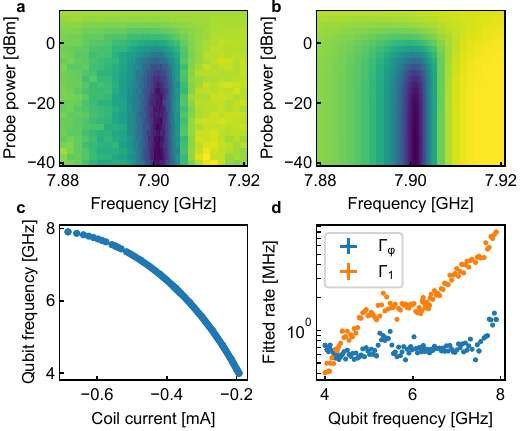} 
			   \caption[power calibration]
			   {
			   	\textbf{Calibration of the frequency dependent power at the TWPA input.}\textbf{ a}, Measured transmittance amplitude through the waveguide-QED device versus VNA probe power and frequency at room temperature. \textbf{b}, Transmittance of the globally fitted model to the data in panel a. \textbf{c}, Fitted qubit frequency versus applied coil current. \textbf{d}, Fitted energy relaxation rate and pure dephasing rate versus fitted qubit frequency. 
				}
			    \label{fig:calibration} 
		\end{figure}

		To calibrate the attenuation between the room-temperature input reference plane and the TWPA input reference plane, we used a frequency-tunable transmon qubit capacitively coupled to a through coplanar waveguide.
		By applying current to an external superconducting coil we can tune the frequency of the transmon qubit, as illustrated in \figref{fig:calibration}c from its sweet-spot at 7.9\,GHz down to 3.5\,GHz where the measurement SNR drops due to the cut-off in HEMT's gain.
		At each coil current, we measured the power-dependent qubit transmittance from the low-power regime, where the qubit behave as an almost perfect mirror that reflects weak indident signals, to the high-power regime, where the qubit was saturated and became transparent.
		A representative colormap of the measured transmission amplitude near the sweet spot is shown in \figref{fig:calibration}a.
		We then fit the complex transmittance \cite{mirhosseini2019cavity,kannan2020generating} as a function of the probe power and frequency for each value of the current through the coil using 4 free parameters: attenuation between the VNA reference plane and the qubit $A$, qubit frequency $\omega_q$, energy decay rate $\Gamma_{1}$ and pure dephasing rate $\Gamma_{\varphi}$:
		\begin{equation}
				\label{eq:transmittance}
				S_{\rm{21}}(\omega) = 1 - \frac{\Gamma_1}{2\Gamma_2}\frac{1 - \frac{i\Delta}{\Gamma_2}}{1 + \left(\frac{\Delta}{\Gamma_2}\right)^2 + \frac{\Omega^2}{\Gamma_1\Gamma_2}},
		\end{equation}
		where $\Delta = \omega - \omega_q$, $\Gamma_2 = \Gamma_1 / 2 + \Gamma_{\varphi}$ is the total dephasing rate, $\Omega^2 = \frac{2AP\Gamma_1}{\hbar\omega_q}$ is the driving strength, $P$ is the probe power at the VNA output and $\hbar$ is the reduced Plank's constant.
		The resulting global fit is plotted in \figref{fig:calibration}b for comparison with the data of panel a.
		The fitted values of the four fitting parameters are plotted in \figref{fig:calibration}c, d as well as in \figref{fig:noise}d.

	\section{Constellation diagram measurement}

	For the IQ-constellation measurements shown in Fig.\ref{fig:power_handling}e--g, the TWPA pump was applied continuously at the operating point used for the power-handling measurements. Square signal pulses with a duration of 2\,$\mu$s were applied to the TWPA input. The pulse amplitude was swept linearly, and the phase was stepped through eight equally spaced values separated by $\pi/4$. For each amplitude--phase setting, $2^{13}$ shots were acquired with a $200\,\mu\mathrm{s}$ inter-pulse delay to build up the corresponding IQ distribution. The transmitted field was demodulated and integrated over a 1.7\,$\mu$s window starting 300\,ns after the onset of the signal pulse. This delayed integration excludes the initial transient response and extracts the IQ point only after the TWPA has reached steady state.

	\begin{figure}[h]
				\centering 
			   \includegraphics[width=\columnwidth]{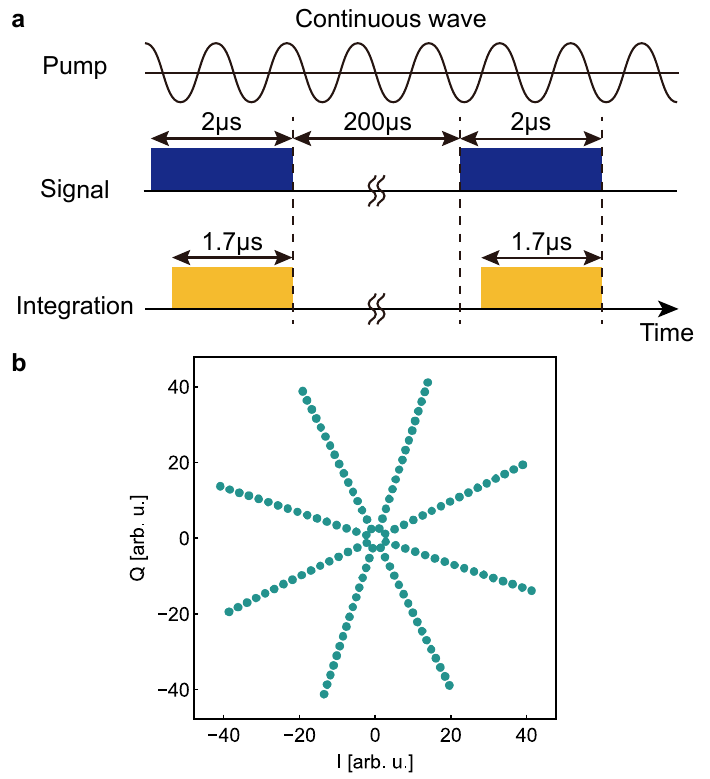} 
			   \caption[pulse sequence]
			   {\textbf{a}, Pulse sequence used for the IQ-constellation measurements. The TWPA pump is applied continuously, while $2\,\mu\mathrm{s}$ square signal pulses are sent to the TWPA input. The IQ integration window starts $300\,\mathrm{ns}$ after the onset of the signal pulse and lasts for $1.7\,\mu\mathrm{s}$. The signal phase is swept in steps of $\pi/4$, and the pulse amplitude is swept linearly. \textbf{b}, IQ constellation diagram measured with the TWPA bypassed, using the same pulse-amplitude and phase sweep as in Fig.~4e but with 20 dB higher signal power. The IQ clusters remain nearly circular, and their centroids follow the programmed pulse amplitude and phase.
				}
			    \label{fig:constellation} 
		\end{figure}

	As a control measurement, we repeated the same pulsed-IQ acquisition through the bypass line with a signal power increased by 20 dB relative to the  signal used in the constellation diagram measurement with TWPA. No IQ deformation or gain saturation of the HEMT or room-temperature amplifier was observed. This control measurement indicates that the deformation of the IQ blobs at high signal power in Fig.~4g originates from TWPA gain compression and nonlinear phase response, rather than saturation of the following amplification or acquisition chain.

	\section{Gain ripple}
	\label{app:gain ripple}
	To assess the sensitivity of the gain ripple to the external microwave environment, we performed an additional diagnostic measurement in an independent cooldown. In that cooldown, two $10\,\mathrm{dB}$ cryogenic attenuators were used before and after the TWPA. The attenuators suppress reflections from the surrounding cryogenic wiring and improve the impedance matching. This measurement was performed in a different dilution refrigerator with a slightly different cryogenic wiring configuration from that used for the  gain, added noise, and power-handling measurements in the main text. Therefore, the data are used only as a qualitative diagnostic of the effect of the external impedance environment, rather than as a calibrated performance comparison.

\begin{figure*}[t]
				\centering 
				\includegraphics[width=\textwidth]{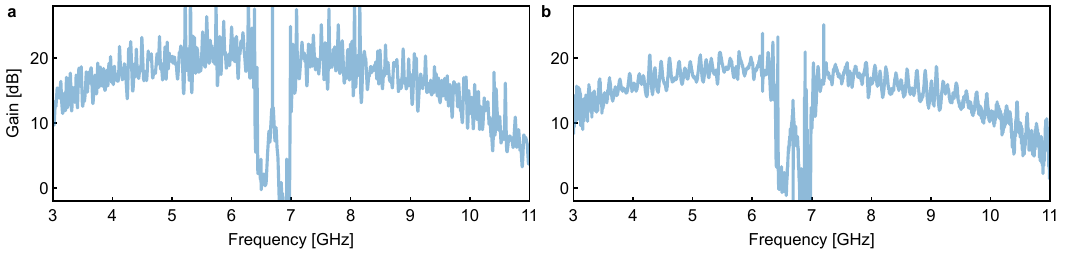} 
				\caption[Gain ripple]
				{\textbf{TWPA gain profile measured with different impedance environment.} \textbf{a}, TWPA gain profile measured with setup shown in Fig.~7, without additional attenuators adjacent to the TWPA.
				\textbf{b}, TWPA gain profile measured in an independent cooldown in a different dilution refrigerator, with the TWPA package placed between two additional $10\,\mathrm{dB}$ cryogenic attenuators. The gain ripple is clearly reduced, while the remaining gain ripple is most likely due to the impedance mismatch from TWPA itself. 
				}
				\label{fig:gain ripple} 
			\end{figure*}

	Figure~\ref{fig:gain ripple} compares the gain profiles measured without and with the additional attenuators. The attenuated configuration shows a reduced gain-ripple amplitude, however no systematic improvement of the maximum stable gain was observed in the attenuated configuration.

	\section{Detailed Fabrication process}

		\subsection{Substrate Preparation}
		We use 4-inch, (100)-oriented, intrinsic float-zone, single-side polished silicon wafers supplied by Topsil.
		These wafers feature high resistivity ($>20$~k$\Omega\cdot$cm), low bow ($<20~\mu$m), and low total thickness variation (TTV $< 5~\mu$m).
		Low TTV is crucial for achieving good lithography uniformity, especially for the phase-matching resonators which have critical dimensions of around 400~nm.

		The wafers are first cleaned with Piranha solution (96\% H$_2$SO$_4$ activated by 30\% H$_2$O$_2$) for 10 minutes at 100$^\circ$C, followed by two rinses in deionized (DI) water and spin drying at room-temperature.
		To remove the native silicon oxide on the surface, the wafers are cleaned with buffered hydrofluoric acid (40\% NH$_4$F and 50\% HF with a volume dilution ratio of 7:1) for 5 minutes.
		This is followed by two DI water rinses and spin drying.
		The wafers are  transferred to the loadlock of the sputtering machine (Pfeiffer SPIDER Sputtering System) within three minutes to prevent re-oxidation.
		A 150~nm niobium film is deposited on the wafers.
		This process has been previously utilized to fabricate transmon qubits with long coherence times \cite{kono2024mechanically}.

		\subsection{Reticle Fabrication and Deep Ultraviolet (DUV) Photolithography} 
		To define the interdigitated capacitors and phase-matching resonators, we employ DUV photolithography.
		The reticle for the DUV lithography is fabricated on a 6-inch Quartz Chrome plate coated with 530~nm of AZ1512HS photoresist.
		The resist is patterned using a direct laser writer (Heidelberg VPG200).
		After exposure, the reticle is processed inside a Hamatech HMR900 mask processor, which includes development with AZ 351B, chromium etching with TechniEtch Cr N$^\circ$1, and resist stripping with AZ 400K.
		Due to our small chip size, we are able to place 9 images on a single 6-inch reticle, with each image corresponding to a design featuring a different phase-matching resonator frequency.

		For the wafer exposure, the substrates are coated with M108Y 400\,nm DUV resist  layer (TEL Clean Track ACT-8).
		They are patterned using a 248~nm DUV stepper (ASML PAS 5500/350C) with a dose of 11~mJ/cm$^2$, a focus of 0~$\mu$m, and a 4$\times$ reduction of the image from the reticle. For the device presented in this work, only the top-left design on the reticle was utilized during the DUV exposure.
		During this step, 8 alignment markers for the subsequent electron-beam (e-beam) and laser-writing lithography steps are simultaneously exposed.
		The wafers then undergo a post-exposure bake at 180$^\circ$C  for 1.5 minutes, followed by development in the TEL Clean Track ACT-8 for 60 seconds.

		\begin{figure}[t]
				\centering 
			    \includegraphics[width=\columnwidth]{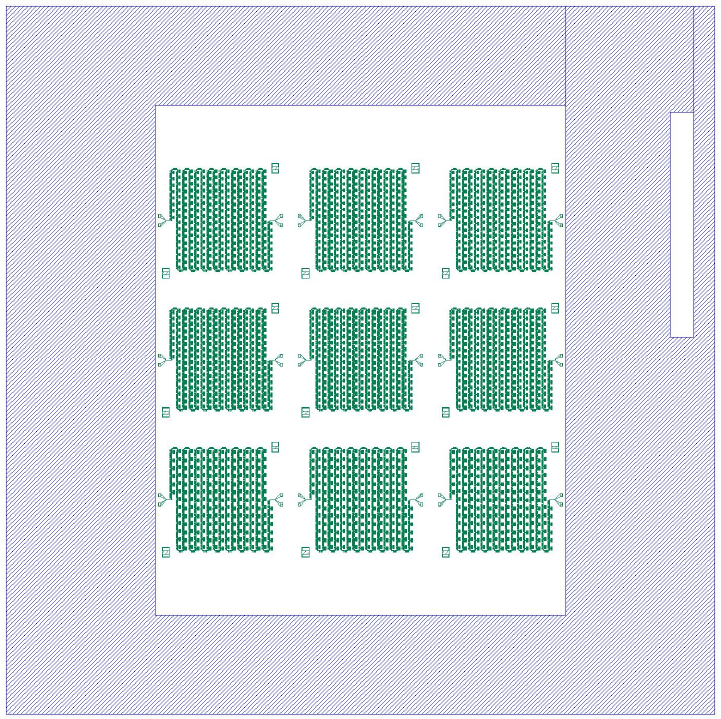} 
			    \caption[Reticle used for DUV lithography]
			    {
			    	A single 6-inch reticle accommodating nine different TWPA designs. The compact chip size enables a single mask to house multiple device variants with differing phase-matching resonator frequencies or dynamic ranges, allowing for selective exposure during DUV lithography.
				}
			    \label{fig:reticle} 
		\end{figure}

		\subsection{Etching and Chip Labeling}
		After DUV exposure, the pattern is transferred to the niobium film using a Cl$_2$/BCl$_3$-based dry etch (Oxford Instruments PlasmaPro100 Cobra), followed by a 5-minute DI water rinse and spin drying.
		All remaining photoresist is fully removed using an electrically inactive N$_2$/5\%H$_2$ mixture (ESI 3511 Downstream Plasma Asher) at 250$^\circ$C for 2 minutes, performed twice. 

		To define the labels for individual chips, 1.5~$\mu$m of AZ ECI 3007 photoresist is spin-coated onto the wafer using an ACS 200 GEN3 system.
		The labels are patterned using a direct laser writer (MLA 150) with a 405~nm laser at a dose of 140~mJ/cm$^2$.
		The photoresist is baked at 110$^\circ$C for 1 minute and developed using AZ 726 MIF (all performed inside the ACS 200 GEN3).
		The label patterns are then transferred to the niobium film using the same Cl$_2$/BCl$_3$-based dry etching process.
		The top thin layer of photoresist, which hardens during the dry etch, is removed using a 2-minute oxygen plasma ash at 200~W (Tepla GiGAbatch).
		The remaining bulk photoresist is stripped using Remover 1165 at 80$^\circ$C.
		Finally, the wafer is cleaned in acetone and isopropanol (IPA) for 3 minutes each and dried with a nitrogen gun.

		\subsection{Junction Fabrication}
		Our Josephson junction fabrication method closely follows the Manhattan-style technique.
		First, approximately 500~nm of MMA EL9 resist is spin-coated onto the wafer, baked at 180$^\circ$C for 5 minutes, and cooled to room temperature.
		Immediately after, a top layer of approximately 500~nm PMMA 495K A8 resist is spin-coated, baked at 180$^\circ$C for 5 minutes, and cooled.
		E-beam lithography is performed using a 100~keV system (Raith EBPG5000+ or EBPG5200 Plus) with a 30~nA beam current, aligned to the markers defined during the prior DUV step.
		The junction regions are exposed with a dose of 1600~$\mu$C/cm$^2$, while the undercuts receive 350~$\mu$C/cm$^2$.
		Following exposure, the wafer is developed in a 1:3 MIBK:IPA solution at room temperature for 2 minutes, dipped in IPA for 1 minute, and dried with a nitrogen gun.

		We utilize double-angle evaporation in a three-chamber Plassys MEB550SL3 system to fabricate the junctions.
		The loadlock is pumped down to $5\times10^{-7}$~Torr before transferring the wafer to the evaporation chamber.
		Prior to the first aluminum deposition, we evaporate titanium at a rate of 0.2~nm/s for 2 minutes with the shutter closed (as a getter).
		We then evaporate 50~nm of aluminum at 0.5~nm/s with the wafer tilted at $\theta=45^\circ$ and no in-plane rotation ($\phi=0^\circ$).
		The tunnel barrier is formed in a dedicated oxidation chamber via dynamic oxidation at 0.03~Torr for 10 minutes.
		The wafer is returned to the evaporation chamber, and a second 110~nm aluminum layer is deposited at 0.5~nm/s with the wafer tilted at $\theta=45^\circ$ and rotated in-plane to $\phi=90^\circ$.
		A final 10-minute oxidation at 15~Torr is performed in the loadlock to form a clean protective oxide layer.
		Lift-off is achieved by soaking the wafer in Remover 1165 at 80$^\circ$C for over 8 hours, using a pipette to flush away the floating aluminum film.
		The wafer is then sonicated in acetone and IPA baths for 5 minutes each and dried with a nitrogen gun.

		To ensure junction reproducibility, the lead widths are fixed at 250~nm.
		We sweep the junction area across 9 different values on the wafer (each increasing by 20\% in area) to guarantee that at least one design yields a characteristic impedance near 50~$\Omega$.

		\subsection{Patches}
		An additional e-beam lithography and lift-off step is used to fabricate patches that establish galvanic contact between the Josephson junctions and the signal line.
		Using the exact same procedure as the junction fabrication, the wafer is coated with the MMA/PMMA bilayer resist stack.
		Exposure is conducted on the 100~keV Raith system with a 100~nA beam current, followed by the standard 1:3 MIBK:IPA development and IPA rinse.

		Before evaporating the aluminum patches, native niobium and aluminum oxides are removed in the Plassys loadlock via argon ion milling for 4.5 minutes (wafer rotating at 5~rpm, no tilt).
		The wafer is moved to the evaporation chamber where titanium is evaporated for 2 minutes (0.2~nm/s)  with the shutter closed, followed by 300~nm of aluminum (wafer rotating at 5~rpm, no tilt).
		A 10-minute oxidation at 15~Torr is performed in the loadlock to passivate the surface.
		Lift-off is performed identically to the junction step.

		\subsection{Airbridges}
		Next, we fabricate airbridges to connect the ground planes and suppress slotline modes.
		A 2~$\mu$m layer of AZ ECI 3027 is spin-coated using the ACS 200 GEN3 and patterned with the MLA 150 direct laser writer.
		The resist is baked at 110$^\circ$C for 90 seconds and developed in AZ 726 MIF.
		The wafer is then baked at 140$^\circ$C to reflow the photoresist, creating the characteristic arc shape of the airbridge scaffold. 

		To remove native niobium oxide, 8 minutes of argon ion milling is performed in the Plassys loadlock (wafer tilted at 45$^\circ$, rotating at 5~rpm).
		In the evaporation chamber, 2 minutes of titanium is evaporated at 0.2~nm/s, followed by 300~nm of aluminum at 0.5~nm/s (wafer rotating at 5~rpm, no tilt).
		The surface is passivated with a 10-minute oxidation at 15~Torr. 

		To pattern the actual airbridge structures, 1.5~$\mu$m of AZ ECI 3007 is spin-coated and exposed using the MLA 150 (with an inverted pattern).
		After development in AZ 726 MIF, a 20-second oxygen plasma ash at 200~W is applied to remove residual descum resist.
		The aluminum film is wet-etched using TechniEtch Al80, a commercial solution of 100\% H$_3$PO$_4$, CH$_3$COOH, and 70\% HNO$_3$ (volume ratio 83:5.5:5.5) at 35$^\circ$C for approximately 2.25 minutes.
		This is followed by two DI water rinses and spin drying.
		Finally, a 2-minute oxygen plasma ash at 200~W is applied to remove any photoresist hardened by the argon milling step.
		Note that the bulk of the 3027 reflow scaffold remains intact after this step.

		\subsection{Dicing and Release}
		Prior to dicing, 1.5~$\mu$m of AZ ECI 3007 photoresist is spin-coated onto the wafer to protect the chip surfaces from debris.
		After dicing, the protective resist and the underlying airbridge scaffold are stripped by soaking the chips in Remover 1165 at 80$^\circ$C for 8 hours.
		This is followed by sequential cleaning in acetone and IPA for 5 minutes each.
		The final chips are gently dried with a nitrogen gun.

	\newpage

	\bibliography{reference}

@article{clerk2010introduction,
  title={Introduction to quantum noise, measurement, and amplification},
  author={Clerk, Aashish A and Devoret, Michel H and Girvin, Steven M and Marquardt, Florian and Schoelkopf, Robert J},
  journal={Reviews of Modern Physics},
  volume={82},
  number={2},
  pages={1155--1208},
  year={2010},
  doi={10.1103/RevModPhys.82.1155},
  publisher={APS}
}

@article{macklin2015near,
  title={A near--quantum-limited {J}osephson traveling-wave parametric amplifier},
  author={Macklin, Chris and O’brien, K and Hover, D and Schwartz, ME and Bolkhovsky, V and Zhang, X and Oliver, WD and Siddiqi, I},
  journal={Science},
  volume={350},
  number={6258},
  pages={307--310},
  year={2015},
  doi={10.1126/science.aaa8525},
  publisher={AAAS}
}

@article{kono2024mechanically,
  title={Mechanically induced correlated errors on superconducting qubits with relaxation times exceeding 0.4 ms},
  author={Kono, Shingo and Pan, Jiahe and Chegnizadeh, Mahdi and Wang, Xuxin and Youssefi, Amir and Scigliuzzo, Marco and Kippenberg, Tobias J},
  journal={Nature Communications},
  volume={15},
  number={1},
  pages={3950},
  year={2024},
  doi={10.1038/s41467-024-48230-3},
  publisher={Nature Publishing Group}
}

@article{castellanos2008amplification,
  title={Amplification and squeezing of quantum noise with a tunable {J}osephson metamaterial},
  author={Castellanos-Beltran, Manuel A and Irwin, KD and Hilton, GC and Vale, LR and Lehnert, KW},
  journal={Nature Physics},
  volume={4},
  number={12},
  pages={929--931},
  year={2008},
  doi={10.1038/nphys1090},
  publisher={Nature Publishing Group}
}

@article{vijay2011observation,
  title={Observation of quantum jumps in a superconducting artificial atom},
  author={Vijay, R and Slichter, DH and Siddiqi, I},
  journal={Physical Review Letters},
  volume={106},
  number={11},
  pages={110502},
  year={2011},
  doi={10.1103/PhysRevLett.106.110502},
  publisher={APS}
}

@article{mutus2014strong,
  title={Strong environmental coupling in a {J}osephson parametric amplifier},
  author={Mutus, Josh Y and White, Theodore C and Barends, Rami and Chen, Yu and Chen, Zijun and Chiaro, Ben and Dunsworth, Andrew and Jeffrey, Evan and Kelly, Julian and Megrant, Anthony and others},
  journal={Applied Physics Letters},
  volume={104},
  number={26},
  pages={263513},
  year={2014},
  doi={10.1063/1.4886408},
  publisher={AIP Publishing}
}

@article{roy2016introduction,
  title={Introduction to parametric amplification of quantum signals with {J}osephson circuits},
  author={Roy, Ananda and Devoret, Michel},
  journal={Comptes Rendus Physique},
  volume={17},
  number={7},
  pages={740--755},
  year={2016},
  doi={10.1016/j.crhy.2016.07.012},
  publisher={Elsevier}
}

@article{eichler2014controlling,
  title={Controlling the dynamic range of a {J}osephson parametric amplifier},
  author={Eichler, Christopher and Wallraff, Andreas},
  journal={EPJ Quantum Technology},
  volume={1},
  number={1},
  pages={2},
  year={2014},
  doi={10.1140/epjqt2},
  publisher={Springer}
}

@article{o2014resonant,
  title={Resonant phase matching of {J}osephson junction traveling wave parametric amplifiers},
  author={O’Brien, Kevin and Macklin, Chris and Siddiqi, Irfan and Zhang, Xiang},
  journal={Physical Review Letters},
  volume={113},
  number={15},
  pages={157001},
  year={2014},
  doi={10.1103/PhysRevLett.113.157001},
  publisher={APS}
}

@article{yaakobi2013parametric,
  title={Parametric amplification in {J}osephson junction embedded transmission lines},
  author={Yaakobi, Oded and Friedland, Lazar and Macklin, Chris and Siddiqi, Irfan},
  journal={Physical Review B},
  volume={87},
  number={14},
  pages={144301},
  year={2013},
  doi={10.1103/PhysRevB.87.144301},
  publisher={APS}
}

@article{aumentado2020superconducting,
  title={Superconducting parametric amplifiers: The state of the art in {J}osephson parametric amplifiers},
  author={Aumentado, Jose},
  journal={IEEE Microwave Magazine},
  volume={21},
  number={8},
  pages={45--59},
  year={2020},
  doi={10.1109/MMM.2020.2993476},
  publisher={IEEE}
}

@article{ho2012wideband,
  title={A wideband, low-noise superconducting amplifier with high dynamic range},
  author={Ho Eom, Byeong and Day, Peter K and LeDuc, Henry G and Zmuidzinas, Jonas},
  journal={Nature Physics},
  volume={8},
  number={8},
  pages={623--627},
  year={2012},
  doi={10.1038/nphys2356},
  publisher={Nature Publishing Group}
}

@article{renberg2024peripheral,
  title={Peripheral circuits for ideal performance of a traveling-wave parametric amplifier},
  author={Renberg Nilsson, Hampus and Shiri, Daryoush and Rehammar, Robert and Fadavi Roudsari, Anita and Delsing, Per},
  journal={Physical Review Applied},
  volume={21},
  number={6},
  pages={064062},
  year={2024},
  doi={10.1103/PhysRevApplied.21.064062},
  publisher={APS}
}

@article{Peng2022,
  title = {Floquet-Mode Traveling-Wave Parametric Amplifiers},
  volume = {3},
  number = {2},
  pages = {020306},
  journal = {PRX Quantum},
  author = {Peng, Kaidong and Naghiloo, Mahdi and Wang, Jennifer and Cunningham, Gregory D. and Ye, Yufeng and O’Brien, Kevin P.},
  year = {2022},
  doi = {10.1103/PRXQuantum.3.020306},
  publisher = {APS}
}

@inproceedings{peng2022xparameter,
  title={X-parameter based design and simulation of Josephson traveling-wave parametric amplifiers for quantum computing applications},
  author={Peng, Kaidong and Poore, Rick and Krantz, Philip and Root, David E. and O'Brien, Kevin P.},
  booktitle={2022 IEEE International Conference on Quantum Computing and Engineering (QCE)},
  pages={331--340},
  year={2022},
  organization={IEEE},
  doi={10.1109/QCE53715.2022.00054}
}

@article{Qiu2023,
  title = {Broadband squeezed microwaves and amplification with a {J}osephson travelling-wave parametric amplifier},
  journal = {Nature Physics},
  volume = {19},
  pages = {706--713},
  author = {Qiu, Jack Y. and Grimsmo, Arne and Peng, Kaidong and Kannan, Bharath and Lienhard, Benjamin and Sung, Youngkyu and Krantz, Philip and Bolkhovsky, Vladimir and Calusine, Greg and Kim, David and others},
  year = {2023},
  doi = {10.1038/s41567-022-01929-w},
  publisher = {Nature Publishing Group}
}

@article{tien1958parametric,
  title={Parametric amplification and frequency mixing in propagating circuits},
  author={Tien, Ping K},
  journal={Journal of Applied Physics},
  volume={29},
  number={9},
  pages={1347--1357},
  year={1958},
  doi={10.1063/1.1723440},
  publisher={AIP}
}

@article{cullen1958travelling,
  title={A travelling-wave parametric amplifier},
  author={Cullen, AL},
  journal={Nature},
  volume={181},
  number={4605},
  pages={332--332},
  year={1958},
  doi={10.1038/181332a0},
  publisher={Nature Publishing Group}
}

@article{vissers2016low,
  title={Low-noise kinetic inductance traveling-wave amplifier using three-wave mixing},
  author={Vissers, Michael R and Erickson, Robert P and Ku, H-S and Vale, Leila and Wu, Xian and Hilton, GC and Pappas, David P},
  journal={Applied Physics Letters},
  volume={108},
  number={1},
  pages={012601},
  year={2016},
  doi={10.1063/1.4937922},
  publisher={AIP Publishing}
}

@inproceedings{zorin2017traveling,
  title={Traveling-wave parametric amplifier based on three-wave mixing in a {J}osephson metamaterial},
  author={Zorin, Alexander B and Khabipov, Marat and Dietel, J and Dolata, R},
  booktitle={2017 16th International Superconductive Electronics Conference (ISEC)},
  pages={1--3},
  year={2017},
  doi={10.1109/ISEC.2017.8314196},
  organization={IEEE}
}

@article{planat2020photonic,
  title={Photonic-crystal {J}osephson traveling-wave parametric amplifier},
  author={Planat, Luca and Ranadive, Arpit and Dassonneville, R{\'e}my and Puertas Mart{\'\i}nez, Javier and L{\'e}ger, S{\'e}bastien and Naud, C{\'e}cile and Buisson, Olivier and Hasch-Guichard, Wiebke and Basko, Denis M and Roch, Nicolas},
  journal={Physical Review X},
  volume={10},
  number={2},
  pages={021021},
  year={2020},
  doi={10.1103/PhysRevX.10.021021},
  publisher={APS}
}

@article{fadavi2023three,
  title={Three-wave mixing traveling-wave parametric amplifier with periodic variation of the circuit parameters},
  author={Fadavi Roudsari, Anita and Shiri, Daryoush and Renberg Nilsson, Hampus and Tancredi, Giovanna and Osman, Amr and Svensson, Ida-Maria and Kudra, Marina and Rommel, Marcus and Bylander, Jonas and Shumeiko, Vitaly and others},
  journal={Applied Physics Letters},
  volume={122},
  number={5},
  pages={052601},
  year={2023},
  doi={10.1063/5.0127690},
  publisher={AIP Publishing}
}

@article{ranadive2022kerr,
  title={Kerr reversal in {J}osephson meta-material and traveling wave parametric amplification},
  author={Ranadive, Arpit and Esposito, Martina and Planat, Luca and Bonet, Edgar and Naud, C{\'e}cile and Buisson, Olivier and Guichard, Wiebke and Roch, Nicolas},
  journal={Nature Communications},
  volume={13},
  number={1},
  pages={1737},
  year={2022},
  doi={10.1038/s41467-022-29375-5},
  publisher={Nature Publishing Group}
}

@article{wang2025high,
  title={High-efficiency, low-loss {F}loquet-mode traveling wave parametric amplifier},
  author={Wang, Jennifer and Peng, Kaidong and Knecht, Jeffrey M and Cunningham, Gregory D and Lombo, Andres E and Yen, Alec and Zaidenberg, Daniela A and Gingras, Michael and Niedzielski, Bethany M and Stickler, Hannah and others},
  journal={arXiv preprint arXiv:2503.11812},
  year={2025},
  doi={10.48550/arXiv.2503.11812}
}

@article{gaydamachenko2025rf,
  title={rf-{SQUID}-based traveling-wave parametric amplifier with input saturation power of- 84 d{B}m across more than one octave in bandwidth},
  author={Gaydamachenko, Victor and Kissling, Christoph and Gr{\"u}nhaupt, Lukas},
  journal={Physical Review Applied},
  volume={23},
  number={6},
  pages={064053},
  year={2025},
  doi={10.1103/1qk4-fzkq},
  publisher={APS}
}

@article{chang2025josephson,
  title={{J}osephson traveling-wave parametric amplifier based on a low-intrinsic-loss lumped-element coplanar waveguide},
  author={Chang, CW Sandbo and Van Loo, Arjan F and Hung, Chih-Chiao and Zhou, Yu and Gnandt, Christian and Tamate, Shuhei and Nakamura, Yasunobu},
  journal={Physical Review Applied},
  volume={24},
  number={4},
  pages={044081},
  year={2025},
  doi={10.1103/qhl6-cz2z},
  publisher={APS}
}

@article{bell2026josephson,
  title={{J}osephson traveling-wave parametric amplifier with inverse {K}err phase matching},
  author={Bell, MT},
  journal={Physical Review Applied},
  volume={25},
  number={1},
  pages={014075},
  year={2026},
  doi={/10.1103/frfk-ty7x},
  publisher={APS}
}

@article{esposito2022observation,
  title={Observation of two-mode squeezing in a traveling wave parametric amplifier},
  author={Esposito, Martina and Ranadive, Arpit and Planat, Luca and Leger, S{\'e}bastien and Fraudet, Dorian and Jouanny, Vincent and Buisson, Olivier and Guichard, Wiebke and Naud, C{\'e}cile and Aumentado, Jos{\'e} and others},
  journal={Physical Review Letters},
  volume={128},
  number={15},
  pages={153603},
  year={2022},
  doi={10.1103/PhysRevLett.128.153603},
  publisher={APS}
}

@article{perelshtein2022broadband,
  title={Broadband continuous-variable entanglement generation using a {K}err-free {J}osephson metamaterial},
  author={Perelshtein, MR and Petrovnin, KV and Vesterinen, Visa and Hamedani Raja, Sina and Lilja, Ilari and Will, Marco and Savin, Alexander and Simbierowicz, Slawomir and Jabdaraghi, Robab Najafi and Lehtinen, Janne S and others},
  journal={Physical Review Applied},
  volume={18},
  number={2},
  pages={024063},
  year={2022},
  doi={10.1103/PhysRevApplied.18.024063},
  publisher={APS}
}

@article{nilsson2024small,
  title={A small footprint travelling-wave parametric amplifier with a high Signal-to-Noise Ratio improvement in a wide band},
  author={Nilsson, Hampus Renberg and Chen, Liangyu and Tancredi, Giovanna and Rehammar, Robert and Shiri, Daryoush and Nilsson, Filip and Osman, Amr and Shumeiko, Vitaly and Delsing, Per},
  journal={arXiv preprint arXiv:2408.16366},
  year={2024},
  doi={10.48550/arXiv.2408.16366}
}

@article{remm2023intermodulation,
  title={Intermodulation distortion in a Josephson traveling-wave parametric amplifier},
  author={Remm, Ants and Krinner, Sebastian and Lacroix, Nathan and Hellings, Christoph and Swiadek, Fran{\c{c}}ois and Norris, Graham J and Eichler, Christopher and Wallraff, Andreas},
  journal={Physical Review Applied},
  volume={20},
  number={3},
  pages={034027},
  year={2023},
  publisher={APS},
  doi={10.1103/PhysRevApplied.20.034027}
}

@article{van2024advanced,
  title={Advanced CMOS manufacturing of superconducting qubits on 300 mm wafers},
  author={Van Damme, J and Massar, S and Acharya, R and Ivanov, Ts and Perez Lozano, D and Canvel, Y and Demarets, M and Vangoidsenhoven, D and Hermans, Y and Lai, JG and others},
  journal={Nature},
  volume={634},
  number={8032},
  pages={74--79},
  year={2024},
  publisher={Nature Publishing Group UK London},
  doi={10.1038/s41586-024-07941-9}
}

@article{ke2026scaffold,
  title={Scaffold-assisted window junctions for superconducting qubit fabrication},
  author={Ke, Chung-Ting and Tsai, Jun-Yi and Chen, Yen-Chun and Xu, Zhen-Wei and Blackwell, Elam and Snyder, Matthew A and Weeden, Spencer and Chen, Peng-Sheng and Lai, Chih-Ming and Sheu, Shyh-Shyuan and others},
  journal={npj Quantum Information},
  year={2026},
  doi={10.1038/s41534-026-01249-4},
  publisher={Nature Publishing Group UK London}
}

@misc{OBrien_JosephsonCircuits_jl,
  author       = {O'Brien, Kevin P.},
  title        = {{JosephsonCircuits.jl}: Frequency-domain harmonic-balance simulation of Josephson circuits},
  howpublished = {\url{https://github.com/kpobrien/JosephsonCircuits.jl}},
}

@article{mirhosseini2019cavity,
  title={Cavity quantum electrodynamics with atom-like mirrors},
  author={Mirhosseini, Mohammad and Kim, Eunjong and Zhang, Xueyue and Sipahigil, Alp and Dieterle, Paul B and Keller, Andrew J and Asenjo-Garcia, Ana and Chang, Darrick E and Painter, Oskar},
  journal={Nature},
  volume={569},
  number={7758},
  pages={692--697},
  year={2019},
  doi={10.1038/s41586-019-1196-1},
  publisher={Nature Publishing Group UK London}
}

@article{kannan2020generating,
  title={Generating spatially entangled itinerant photons with waveguide quantum electrodynamics},
  author={Kannan, Bharath and Campbell, Daniel L and Vasconcelos, Francisca and Winik, Roni and Kim, DK and Kjaergaard, Morten and Krantz, Philip and Melville, Alexander and Niedzielski, Bethany M and Yoder, JL and others},
  journal={Science advances},
  volume={6},
  number={41},
  pages={eabb8780},
  year={2020},
  doi={10.1126/sciadv.abb8780},
  publisher={American Association for the Advancement of Science}
}

\end{document}